\documentclass[twocolumn,showpacs,preprintnumbers,prb]{revtex4}% Physical Review B

\usepackage{graphicx}% Include figure files
\usepackage{dcolumn}% Align table columns on decimal point
\usepackage{bm}% bold math

\begin{document}

%\preprint{APS/123-QED}

\title{Intrinsic room temperature ferromagnetism in Co-implanted ZnO}

\author{Numan Akdogan}
 \altaffiliation{Author to whom correspondence should be addressed. E-mail address:
numan.akdogan@ruhr-uni-bochum.de \\
Present address: Department of Physics, Gebze Institute of
Technology, Gebze, 41400, Kocaeli, Turkey}
\author{Alexei Nefedov}
 \altaffiliation{Present address: Lehrstuhl f\"{u}r Physikalische Chemie I, Ruhr-Universit\"{a}t Bochum, D-44780 Bochum, Germany}
\author{Kurt Westerholt}
\author{Hartmut Zabel}
\affiliation{Institut f\"{u}r
Experimentalphysik/Festk\"{o}rperphysik, Ruhr-Universit\"{a}t
Bochum, D-44780 Bochum, Germany}

\author{Hans-Werner Becker}
 \affiliation{Institut f\"{u}r Physik mit Ionenstrahlen, Ruhr-Universit\"{a}t Bochum, D-44780 Bochum, Germany}

\author{Christoph Somsen}
 \affiliation{Institut f\"{u}r Werkstoffe, Ruhr-Universit\"{a}t Bochum, D-44780 Bochum, Germany}

\author{Rustam Khaibullin}
\author{Lenar Tagirov}
 \altaffiliation{Permanent address: Kazan State University, 420008 Kazan, Russia}
\affiliation{Kazan Physical-Technical Institute of RAS, 420029
Kazan, Russia}

\date{\today}% It is always \today, today,

\begin{abstract}
We report on the structural and magnetic properties of a
cobalt-implanted ZnO film grown on a sapphire substrate. X-ray
diffraction and transmission electron microscopy reveal the presence
of a (10\={1}0)-oriented hexagonal Co phase in the Al$_2$O$_3$
sapphire substrate, but not in the ZnO film. Co clusters, with a
diameter of is about 5-6 nm, form a Co rich layer in the substrate
close to the ZnO/Al$_2$O$_3$ interface. Magnetization measurements
indicate that there exist two different magnetic phases in the
implanted region. One originates from the Co clusters in
Al$_2$O$_3$, the other one belongs to a homogeneous ferromagnetic
phase with a ferromagnetic Curie temperature far above room
temperature and can be attributed to Co substitution on Zn sites in
the ZnO layer. We have observed magnetic dichroism at the Co
$\emph{L}_{2,3}$ and O \emph{K} edges at room temperature as well as
the multiplet structure in x-ray absorption spectra around the Co
$\emph{L}_3$ edge, supporting the intrinsic nature of the observed
ferromagnetism in Co-implanted ZnO film. The magnetic moment per
substituted cobalt is found about 2.81 $\mu_B$ which is very close
to the theoretical expected value of 3 $\mu_B/Co$ for Co 2+ in its
high spin state.
\end{abstract}


\maketitle

\section{\label{sec:level1}Introduction}

Spintronics, a short notation for spin-based electronics, is a new
research area which tries to exploit the spin of electrons in
addition to their charge in semiconductor devices. The basic idea is
to combine the characteristics of existing magnetic devices with
semiconductor devices in order to realize a new generation of
devices that are smaller, energy efficient, and faster than
presently available.
\cite{FeynmanFP86,DattaAPL90,PrinzSci98,OhnoAPL98,OestreichN99,OhnoN99,WolfSci01,SarmaAS01,JonkerAPL01}
The key requirement in the development of such devices is an
efficient injection, transfer and detection of spin-polarized
currents. Due to the well known problem of resistance mismatch at
metal/semiconductor interfaces, hindering an effective spin
injection, \cite{SchmidtPRB00} much interest is now concentrating on
the development of room-temperature ferromagnetic semiconductors.

Diluted magnetic semiconductors (DMS) refer to the fact that some
fraction of atoms in a non-magnetic semiconductor is replaced by
magnetic ions. DMS are promising candidates for spintronic
applications at ambient temperatures, provided that their Curie
temperature (T$_C$) is far enough above room temperature. Therefore,
a number of different semiconductor hosts have been investigated to
test their magnetic properties. In the past most attention has been
paid to (Ga, Mn)As
\cite{OhnoAPL96,AndoJAP98,BeschotenPRL99,GrandidierAPL00,SadowskiAPL01,PotashnikAPL01,SchottAPL01}
and (In, Mn)As
\cite{MunekataPRL89,MunekataAPL93,SooPRB96,KoshiharaPRL97,AkaiPRL98,OiwaAPL01}
DMS systems. However their reported highest Curie temperatures are
only around 170 K for (Ga, Mn)As \cite{OhnoSci98,OhnoJVS00} and 35 K
for (In, Mn)As. \cite{OhnoJVS00,OhnoJSAP02} Therefore, there is a
large incentive for developing new DMS materials with much higher
Curie temperatures. In particular, the calculations of Dietl
\emph{et al.} \cite{DietlSci00} were the first to indicate that
Mn-doped ZnO could exhibit ferromagnetism above room temperature.
Later, Sato \emph{et al.} have also investigated ZnO-based DMS by
\emph{ab initio} electronic structure calculations and reported
ferromagnetic ordering of 3\emph{d} transition metal ions in ZnO.
\cite{SatoJJAP00,SatoJJAP01} These theoretical predictions initiated
an outburst of experimental activities of TM-doped ZnO.
\cite{UedaAPL01,PeartonMSE03,PrellierJPCM03,TuanPRB04,PeartonSST04,PeartonJVST04,
LiuJMS05,JanischJPCM05,LiuJAP06,PacuskiPRB06,ZhangJPCM07,BehanPRL08}
Actually, some of these studies indeed claim ferromagnetic signals
above room temperature. However, the origin of ferromagnetism in
this system is still under debate. The main unresolved question is
whether the observed ferromagnetism originates from uniformly
distributed TM elements in the ZnO host matrix or whether it is due
to the precipitation of metallic ferromagnetic clusters. Only a
detailed analysis of the magnetic properties by using a combination
of powerful material characterization techniques might help to
unravel the origin of ferromagnetism in doped ZnO.

The main aim of the present article is to present a corresponding
detailed study of the structural and magnetic properties of a
Co-implanted ZnO film grown on a sapphire substrate. Rutherford
backscattering spectroscopy (RBS), X-ray diffraction (XRD), and high
resolution transmission electron microscopy (TEM) techniques were
used to determine the depth distribution of the implanted cobalt
ions and to detect the formation of possible secondary phases such
as metallic cobalt clusters in the implanted region. To determine
whether the implanted cobalt ions are in the Co$^{2+}$ oxidation
state or exhibit pure metallic behavior, x-ray absorption
spectroscopy (XAS) experiments were also performed. The magnetic
properties were characterized by using the magneto-optical Kerr
effect (MOKE), a superconducting quantum interference device (SQUID)
magnetometer, as well as x-ray resonant magnetic scattering (XRMS)
techniques.

\section{\label{sec:level1}Sample Preparation}

About 350~{\AA} thick ZnO film were grown on $10\times10$ mm$^{2}$
epi-polished single-crystalline Al$_2$O$_3$ ($11\overline{2}0$)
substrate by RF (13.56 MHz) sputtering of a ZnO
target.\cite{AyASS03} The sputtering was carried out in an
atmosphere of $5\times10^{-3}$ mbar pure Ar ($99.999 \%$) with a
substrate temperature of $500^\circ C$. In order to increase the
quality of ZnO film, we have carried out post-growth annealing in an
oxygen atmosphere with a partial pressure of up to 2000 mbar and a
temperature of $800^\circ C$. After annealing, the ZnO sample was
implanted in the ILU-3 ion accelerator (Kazan Physical-Technical
Institute of Russian Academy of Science) by using 40 keV Co$^+$ ions
with an ion current density of $8 \mu A\cdot cm^{-2}$. The sample
holder was cooled by flowing water during the implantation to
prevent the sample from overheating. The implantation dose was
$1.50\times10^{17} ions\cdot cm^{-2}$.

\section{\label{sec:level1}Experimental Results}

\subsection{\label{sec:level2}Structural Properties}

\begin{figure}[!h]
\includegraphics[width=0.4\textwidth]{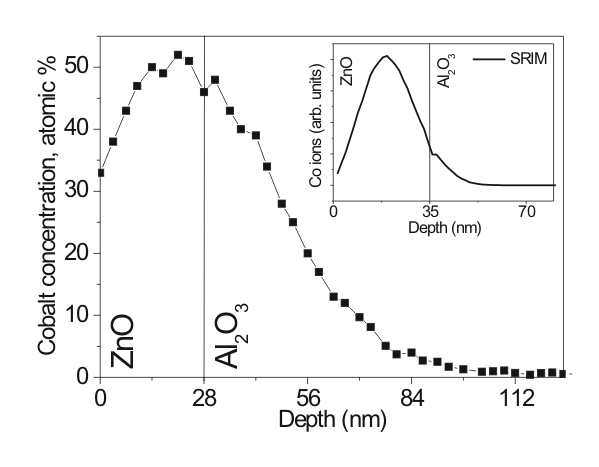}
\caption{\label{RBS-SRIM-ZnO} Depth dependence of the cobalt
concentration in ZnO/Al$_2$O$_3$ implanted with Co ions with a dose
of $1.50\times10^{17} ions\cdot cm^{-2}$. The inset shows the
calculated SRIM profile without taking into account ion sputtering
effects.}
\end{figure}

The depth dependence of the cobalt concentration in Co-implanted
ZnO/Al$_2$O$_3$ film was investigated using RBS technique at the
Dynamic Tandem Laboratory (DTL) at Ruhr-Universit\"{a}t Bochum. The
RBS data shows both a maximum of cobalt concentration (about 50
at.\%) located close to the ZnO/Al$_2$O$_3$ interface and an
extended inward tail due to cobalt diffusion into the volume of the
Al$_2$O$_3$ substrate (Fig.~\ref{RBS-SRIM-ZnO}). It is also observed
that after ion implantation the thickness of the ZnO layer has
decreased from originally 35 nm to 28 nm. According to the SRIM
algorithm,\cite{ZieglerPP85} the average implanted depth of 40 keV
Co ions in ZnO/Al$_2$O$_3$ is about 20.4 nm with a straggling of 9.6
nm in the Gaussian-like depth distribution (inset in
Fig.~\ref{RBS-SRIM-ZnO}). However, because of the surface
sputtering, ion mixing and heating of the implanted region by the
ion beam, there is a redistribution of the implanted cobalt compared
to the calculated profile.

\begin{figure}[!h]
\includegraphics[width=0.4\textwidth]{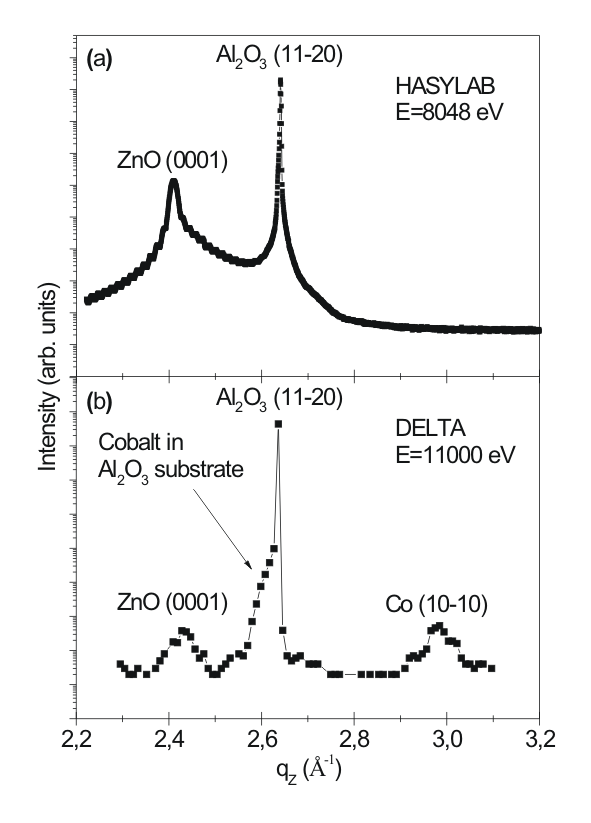}
\caption{\label{Xrd-ZnO} High angle Bragg scans of the ZnO film
before (a) and after (b) cobalt ion implantation.}
\end{figure}

The high-angle XRD experiments provide information on the structural
coherence of the films and in our case also give us a chance to
detect possible additional phases in the sample after ion
implantation. Fig.~\ref{Xrd-ZnO} shows a high angle Bragg scans of
the ZnO film before (a) and after (b) cobalt implantation. The data
were taken using synchrotron radiation at the "Hamburg Synchrotron
Radiation Laboratory" (HASYLAB) (Fig.~\ref{Xrd-ZnO}(a)) and at the
"Dortmund Electron Accelerator" (DELTA) (Fig.~\ref{Xrd-ZnO}(b)) with
an energy of E=8048 eV and E=11000 eV, respectively. X-ray
diffraction measurements yielded evidence for the
($10\overline{1}0$) reflection of the Co hcp structure as is clearly
seen on the right side of the sapphire substrate peak
(Fig.~\ref{Xrd-ZnO}(b)). The heavy ion bombarding also causes a
reduction of intensity of the ZnO (0001) peak. In addition to this,
after implantation we observed that the ZnO (0001) peak is shifted
to higher angles. This is due to the shrinking of the ZnO lattice
caused by the substitution of cobalt ions in the ZnO matrix. After
implantation a tail (shown by an arrow in Fig.~\ref{Xrd-ZnO})
appears around the main peak of Al$_2$O$_3$ ($11\overline{2}0$)
reflection which is not observed before implantation. This tail
likely reflects the lattice expansion of the sapphire substrate upon
Co implantation.

\begin{figure}[!h]
\centering
\includegraphics[width=0.4\textwidth]{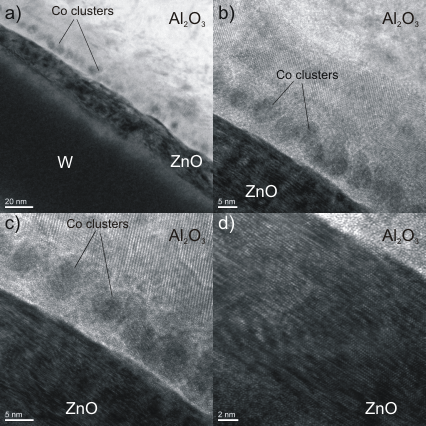}
\caption{\label{ZnO-Tem-S6} TEM images of Co-implanted
ZnO/Al$_2$O$_3$ film. Cobalt clusters are clearly seen in
Al$_2$O$_3$ substrate.}
\end{figure}

In order to further investigate both the presence of metallic cobalt
clusters and the damage of the sapphire substrate, high resolution
cross sectional TEM measurements were performed. The sample
preparation for TEM measurements is done by using focused ion beam
(FIB) technique. To prevent charging effects, the sample surface was
covered by a tungsten (W) film and then a very small cross sectional
piece of the implanted sample was cut by using FIB.
Fig.~\ref{ZnO-Tem-S6} presents TEM images of the ZnO sample with an
increasing magnification from 20 nm to 2 nm. In the first image
(Fig.~\ref{ZnO-Tem-S6}(a)), a general overview of the
ZnO/Al$_2$O$_3$ sample is shown. The cobalt clusters can be seen in
the sapphire substrate located close to the ZnO/Al$_2$O$_3$
interface. The clusters most likely form because of a segregation of
Co in the Al$_2$O$_3$ substrate. Clustering occurs in a Al$_2$O$_3$
at an annealing temperature of $900^\circ C$.\cite{MoraweJAP95}
Thus, obviously ion bombardment heats up the sample locally to this
temperature. Fig.~\ref{ZnO-Tem-S6}(b) and (c) focuses on the
ZnO/Al$_2$O$_3$ interface. These images reveal that the cobalt
clusters have a size of about 5-6 nm and that they nearly touch each
other. Further information from these images is the deformation of
the Al$_2$O$_3$ crystal structure close to the ZnO/Al$_2$O$_3$
interface. This results in a lattice expansion of the substrate
which is in agreement with the XRD results shown in
Fig.~\ref{Xrd-ZnO}. However, far from the interface the structure of
Al$_2$O$_3$ is preserved and one can see nicely the atomic rows of
Al$_2$O$_3$ presented in Fig.~\ref{ZnO-Tem-S6}(c).
Fig.~\ref{ZnO-Tem-S6}(d) shows the ZnO layer with a magnification of
2 nm. Even after heavily ion bombardment, ZnO still has a good
arrangement of atomic rows. Moreover, any distinct clusters cannot
be observed in this region.

\subsection{\label{sec:level2} Magnetic Properties}

\subsubsection{\label{sec:level3} MOKE and SQUID measurements}

\begin{figure}[!h]
\centering
\includegraphics[width=0.4\textwidth]{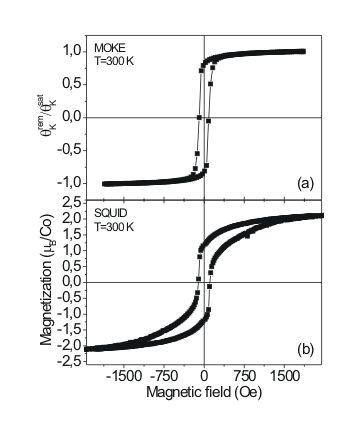}
\caption{\label{ZnO-Moke-Squid-Hys-S6} The MOKE (a) and SQUID (b)
hysteresis curves of Co-implanted ZnO film.}
\end{figure}

Next we discuss the magnetic properties of the Co-implanted ZnO
film. In Fig.~\ref{ZnO-Moke-Squid-Hys-S6}(a) we show the hysteresis
loop of Co-doped ZnO film. This hysteresis was recorded at room
temperature using a high-resolution MOKE setup
\cite{ZeidlerPRB96,Schmitte-diss,Westphalen-diss} in the
longitudinal configuration with s-polarized light. The MOKE data
clearly indicate that after cobalt implantation, non-magnetic ZnO
becomes ferromagnetic at room temperature with a large remanent
magnetization.

Since the MOKE technique is only sensitive to the magnetization of
thin layers close to the surface (20-30 nm penetration depth), $M-H$
measurements have also been carried out using a Quantum Design MPMS
XL SQUID magnetometer. Fig.~\ref{ZnO-Moke-Squid-Hys-S6}(b) presents
the SQUID hysteresis loop of the sample after subtraction of a
diamagnetic contribution from the sapphire substrate. The coercive
field of this hysteresis is more or less the same as the one
measured by MOKE technique. However, some additional contributions
appear and the magnetization saturates at considerably higher
fields.

\subsubsection{\label{sec:level3}XRMS and XAS measurements}

In order to study in detail the observed ferromagnetic behavior, the
magnetic properties of the Co-implanted ZnO film were investigated
using the XRMS and XAS techniques.

XRMS has proven to be a highly effective method for the analysis of
the magnetic properties of buried layers and interfaces, including
their depth dependence. \cite{TonnerrePRL95,LaanCOSSMS06} Moreover,
if the photon energy is fixed close to the energy of the
corresponding absorption edges, element specific hysteresis loops
can be measured. \cite{KortrightNIMB03} Since there are three
elements in the Co-doped ZnO film, the analysis can be carried out
separately for Co, O and Zn.

The XRMS experiments were performed using the ALICE diffractometer
\cite{GrabisRSI03} at the undulator beamline UE56/1-PGM at BESSY II
(Berlin, Germany). The diffractometer comprises a two-circle
goniometer and works in horizontal scattering geometry. A magnetic
field can be applied in the scattering plane along the sample
surface either parallel or antiparallel to the photon helicity,
which corresponds to the longitudinal magneto-optical Kerr effect
(L-MOKE) geometry. The maximum field of $\pm2700 Oe$ was high enough
to fully saturate the sample. The magnetic contribution to the
scattered intensity was always measured by reversing the magnetic
field at fixed photon helicity.

\begin{figure}[!h]
\centering
\includegraphics[width=0.4\textwidth]{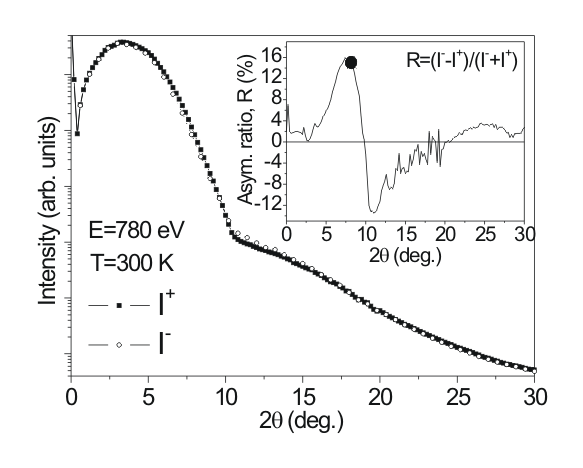}
\caption{\label{ZnO-Xrms-Refl-S6} Reflectivity scans of the sample
taken at the Co $\emph{L}_3$ edge (E=780 eV) with a magnetic field
applied in the sample plane parallel ($I^+$, solid line) and
antiparallel ($I^-$, open circles) to the photon helicity. The inset
shows the asymmetry ratio ($R$) as a function of angle.}
\end{figure}

Fig.~\ref{ZnO-Xrms-Refl-S6} shows the specular reflectivities
measured at the Co $\emph{L}_3$ edge (E=780 eV) in magnetic
saturation. The measurements were taken at room temperature and with
a magnetic field applied in the sample plane parallel ($I^+$, solid
line) and antiparallel ($I^-$, open circles) to the photon helicity.
Due to the high surface roughness no Kiessing fringes are observed
in the reflectivity curves. Nevertheless, the splitting of the two
curves is clearly seen in Fig.~\ref{ZnO-Xrms-Refl-S6}. The inset in
Fig.~\ref{ZnO-Xrms-Refl-S6} presents the angular dependence of the
asymmetry ratio ($R=(I^+-I^-)/(I^++I^-)$) to show how the magnetic
signal varies. As a compromise between high scattering intensity and
high magnetic sensitivity for the investigation of the magnetic
properties at the Co \emph{L} edges, the scattering angle was fixed
at the position of $2\theta=8.2^\circ$ (the angle of incidence
$\theta=4.1^\circ$), shown by a black circle in the inset in
Fig.~\ref{ZnO-Xrms-Refl-S6}. For measurements at the O \emph{K} edge
(E$\sim$530 eV) the scattering angle was fixed at
$2\theta=12^\circ$, which corresponds to the same scattering vector
in the reciprocal space.

\begin{figure}[!h]
\centering
\includegraphics[width=0.4\textwidth]{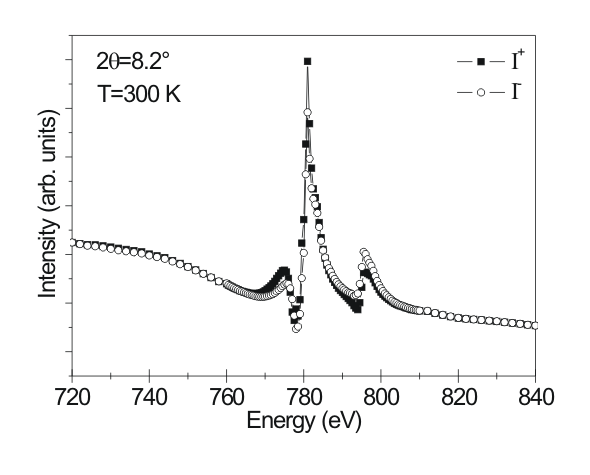}
\caption{\label{ZnO-Xrms-Co-S6} Energy dependence of scattering
intensities at the Co $\emph{L}_{2,3}$ edges measured at room
temperature.}
\end{figure}

\begin{figure}[!h]
\centering
\includegraphics[width=0.4\textwidth]{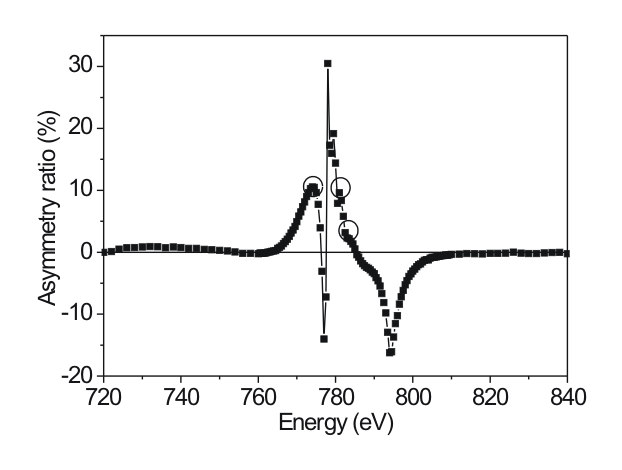}
\caption{\label{ZnO-Xrms-Asym-Co-ZnO} The asymmetry ratio taken at
the Co $\emph{L}_{2,3}$ edges. The circles show the energies where
different magnetic hysteresis curves recorded.}
\end{figure}

The energy dependence of the scattered intensity around the Co
$\emph{L}_{2,3}$ edges measured in positive (solid line) and
negative (open circles) saturation fields is shown in
Fig.~\ref{ZnO-Xrms-Co-S6}. Since the magnetic contribution to the
resonant scattering can best be visualized by plotting the asymmetry
ratio, in Fig.~\ref{ZnO-Xrms-Asym-Co-ZnO} we present the asymmetry
ratio at the Co $\emph{L}_{2,3}$ edges. The asymmetry ratio shows a
strong ferromagnetic signal of up to 30 \%. The fine structure of
the Co $\emph{L}_3$ peak in Fig.~\ref{ZnO-Xrms-Co-S6} is typical for
oxidized cobalt which has been observed before for CoO by Regan
\emph{et al.} \cite{ReganPRB01} They also showed that in the case of
metallic cobalt the Co $\emph{L}_3$ peak consists of mainly one
single component.

\begin{figure}[!h]
\centering
\includegraphics[width=0.4\textwidth]{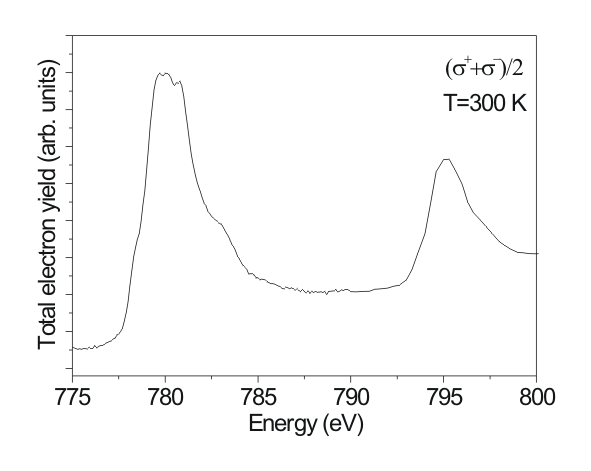}
\caption{\label{ZnO-XAS-S6} X-ray absorption spectra measured at the
Co $L_{2,3}$ edges by using TEY method. $\sigma^+$ and $\sigma^-$
denote the right and left circularly polarized light, respectively.}
\end{figure}

Recently, Kobayashi \emph{et al.} \cite{KobayashiPRB05} reported
that for a Co$^{2+}$ oxidation state in ZnO, the XAS spectra
exhibits a multiplet fine structure around the Co \emph{L}$_3$ edge.
To check whether this behavior is also present in our sample, XAS
experiments were carried out at the undulator beamline UE52-SGM at
BESSY II using the ALICE diffractometer. The absorption data were
taken by the total electron yield (TEY) method, i.e. by measuring
the sample drain current. Since the excited electron trajectories
are affected by the external magnetic field, the XAS spectra were
taken with fixed photon helicity at remanence. The angle of
incidence was chosen to be $4.1^\circ$ with respect to the surface.
The spectra were normalized to the incoming photon flux.
Fig.~\ref{ZnO-XAS-S6} shows the averaged x-ray absorption spectra
($\sigma^+$+$\sigma^-$)/2 at the Co $L_{2,3}$ edges. The XAS
spectrum clearly shows a multiplet structure at the $L_3$ edge which
is similar to that observed before for Co-doped ZnO by Kobayashi
\emph{et al.} This multiplet structure is a clear indication of the
presence of oxidized cobalt in this sample.

\begin{figure}[!h]
\centering
\includegraphics[width=0.4\textwidth]{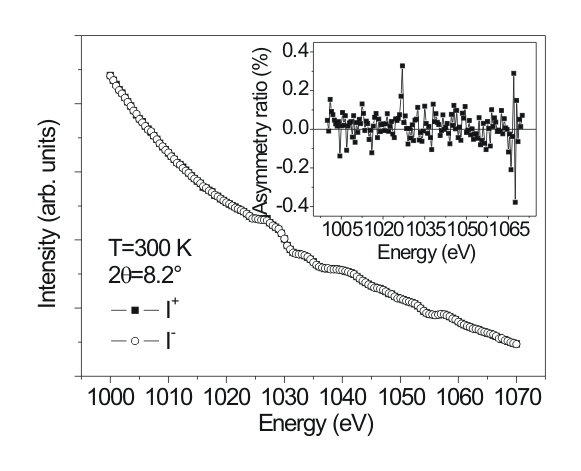}
\caption{\label{ZnO-Xrms-Zn-S6} Energy scan of the reflected
intensity taken at the Zn $\emph{L}_3$ edge. The inset shows the
asymmetry ratio }
\end{figure}

\begin{figure}[!h]
\centering
\includegraphics[width=0.4\textwidth]{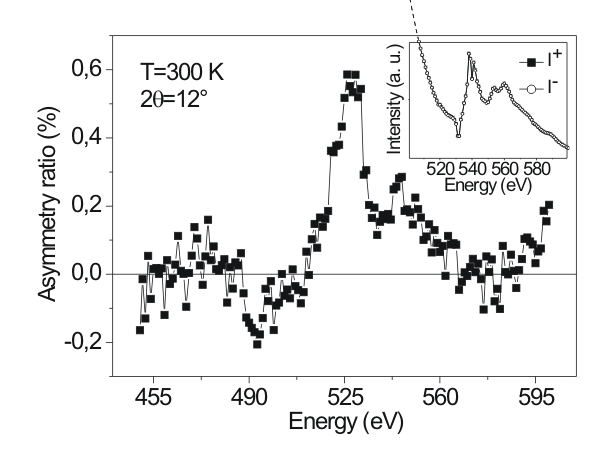}
\caption{\label{ZnO-Xrms-Asym-O-S6} The asymmetry ratio taken at the
O \emph{K} edge. The inset shows the energy scan of the reflected
intensity at the O \emph{K} edge.}
\end{figure}

The magnetic signal around the Zn $\emph{L}_3$ (E=1021.8 eV) and the
O \emph{K} edges were also investigated. Within the sensitivity
limit no magnetic signal could be recorded for Zn
(Fig.~\ref{ZnO-Xrms-Zn-S6}). However, a clear magnetic signal was
observed at the O \emph{K} edge. The asymmetry ratio measured at the
O \emph{K} edge is presented in Fig.~\ref{ZnO-Xrms-Asym-O-S6}. Note
that the maximum in the asymmetry ratio of oxygen is much smaller
(by roughly a factor of fifty) than the asymmetry ratio of cobalt
shown in Fig.~\ref{ZnO-Xrms-Asym-Co-ZnO}.

\begin{figure}[!h]
\centering
\includegraphics[width=0.4\textwidth]{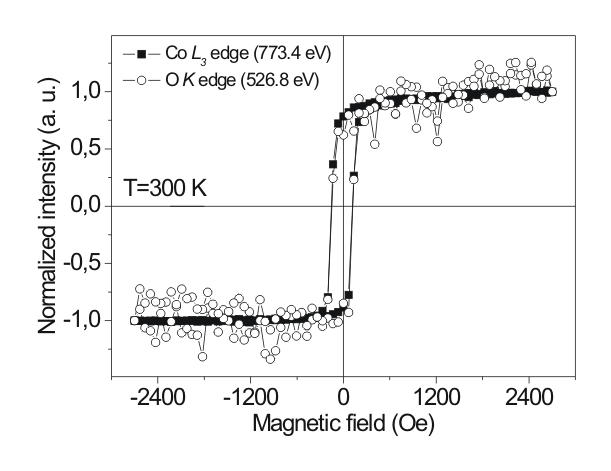}
\caption{\label{ZnO-Xrms-Hys-Co-O-S6} Normalized hysteresis curves
measured at the Co $\emph{L}_3$ (closed symbols) and O \emph{K}
(open symbols) edges.}
\end{figure}

In Fig.~\ref{ZnO-Xrms-Hys-Co-O-S6} we compare the magnetic
hysteresis curves recorded at the Co $\emph{L}_3$ (773.4 eV) and O
\emph{K} (526.8 eV) edges. The shape and the coercive field of the
hysteresis curves are the same, but the intensity is much lower for
the O \emph{K} edge. This is a clear indication for a spin
polarization of oxygen atoms in the ZnO host matrix.

\begin{figure}[!h]
\centering
\includegraphics[width=0.4\textwidth]{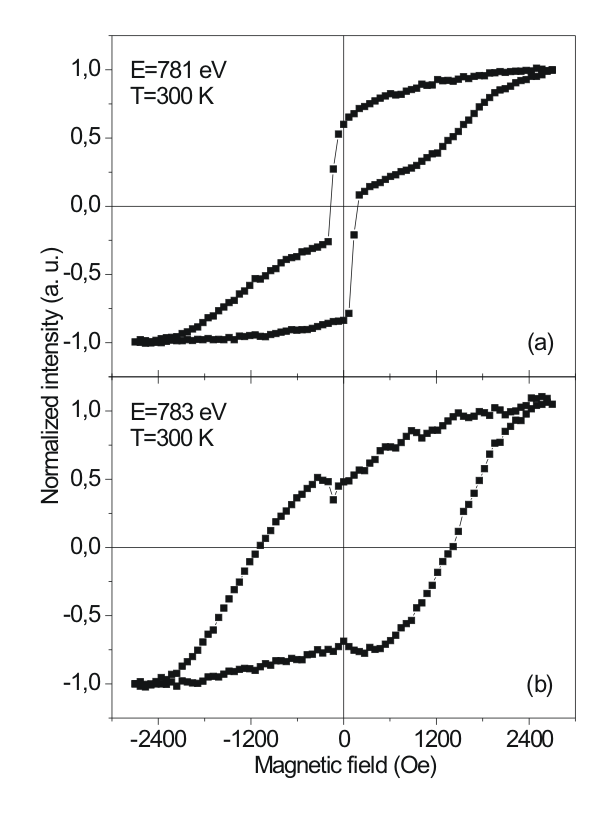}
\caption{\label{ZnO-Xrms-Hys-Diff-S6} (a) Hysteresis loop measured
at the energy of 781 eV shows the superposition of two different
phases of cobalt in the host material. (b) Hysteresis loop taken at
the energy of 783 eV.}
\end{figure}

Since an additional magnetic contribution is observed in the SQUID
hysteresis curve (Fig.~\ref{ZnO-Moke-Squid-Hys-S6}(b)), it was
checked by using XRMS technique whether this feature becomes also
visible. For this reason, several hysteresis loops were recorded at
different photon energies and a systematic change of the hysteresis
loop shape was observed with changing photon energy. Here, only
three hysteresis curves shown by open red circles in
Fig.~\ref{ZnO-Xrms-Asym-Co-ZnO} are presented. The shape of the
hysteresis curve taken at 773.4 eV (closed symbols in
Fig.~\ref{ZnO-Xrms-Hys-Co-O-S6}) is practically the same as the one
measured by MOKE (Fig.~\ref{ZnO-Moke-Squid-Hys-S6}(a)). However,
when the incoming photon energy is increased to 781 eV and 783 eV,
two different hysteresis loops are observed.
Fig.~\ref{ZnO-Xrms-Hys-Diff-S6}(a) presents the hysteresis curve
recorded at 781 eV. At this energy the hysteresis curve has two
components one with a small and one with a large coercive field and
it is similar to the SQUID hysteresis
(Fig.~\ref{ZnO-Moke-Squid-Hys-S6}(b)). At higher energies the low
coercive field component vanishes and at the energy of 783 eV
(Fig.~\ref{ZnO-Xrms-Hys-Diff-S6}(b)) the hysteresis curve consists
of practically only one component with the large coercive field.
This large coercive field component originates from the strong
interaction between cobalt clusters in the sapphire substrate and it
is present even at room temperature.

The reason for the observation of different hysteresis curves using
XRMS technique can be interpreted by the change of the optical
parameters as a function of incident photon energy. Depending on the
energy deviation from the $\emph{L}_{2,3}$ resonance condition, both
types of magnetic hysteresis can be detected. For an energy of 773.4
eV, which is very close to the $\emph{L}_3$ resonance energy, the
contribution to the scattering intensity from the phase with the
large coercive field vanishes, resulting in a hysteresis loop
(closed symbols in Fig.~\ref{ZnO-Xrms-Hys-Co-O-S6}) which is similar
to the one measured by MOKE (Fig.~\ref{ZnO-Moke-Squid-Hys-S6}(a)).
However, at the photon energy of 781 eV , at the resonance
condition, both phases contribute to the scattering intensity. The
hysteresis taken at this energy (Fig.~\ref{ZnO-Xrms-Hys-Diff-S6}(a))
presents the superposition of two phases of cobalt in the host
material and it is similar to the hysteresis measured with the SQUID
magnetometer (Fig.~\ref{ZnO-Moke-Squid-Hys-S6}(b)). The hysteresis
loop recorded at the energy of 783 eV
(Fig.~\ref{ZnO-Xrms-Hys-Diff-S6}(b)) is representative for the
metallic phase of cobalt in Al$_2$O$_3$ with a large coercive field,
whereas at this energy the contribution to the scattering intensity
from the small coercive field component nearly vanishes.

\subsubsection{\label{sec:level3} Magnetic moment per substituted Co atom and estimation for T$_C$}

To calculate the magnetic moment per substituted Co atom we used the
RBS data and SQUID hysteresis presented in Figs.~\ref{RBS-SRIM-ZnO}
and ~\ref{ZnO-Moke-Squid-Hys-S6}. First we determined the percentage
of cobalt atoms located within ZnO layer from the RBS data.
Integrating the area under the curve in Fig.~\ref{RBS-SRIM-ZnO}, we
estimate that 48.5 per cent of the implanted cobalt atoms are
contained inside the ZnO layer. Secondly, we assumed that all cobalt
atoms in the ZnO layer are substituted in the ZnO lattice. Indeed,
the XAS data presented in Fig.~\ref{ZnO-XAS-S6} provide clear
evidence for substitutional cobalt in the implanted matrix and
furthermore we could not find any indication for clustering in the
ZnO layer in the TEM pictures. We also assumed that all cobalt atoms
in the Al$_2$O$_3$ substrate are in the cluster phase with a
magnetic moment of 1.6 $\mu_B$ per cobalt atom, as expected for
metallic cobalt. Using the average magnetic moment value from the
SQUID data ($\mu$=2.19 $\mu_B/Co$), finally we calculate $\mu$=2.81
$\mu_B$ per substituted cobalt atom in the ZnO layer. This value is
very close to the magnetic moment of Co 2+ in its high spin state
($\mu$=3 $\mu_B/Co$). \cite{JanischJPCM05,PattersonPRB06}

\begin{figure}[!h]
\centering
\includegraphics[width=0.4\textwidth]{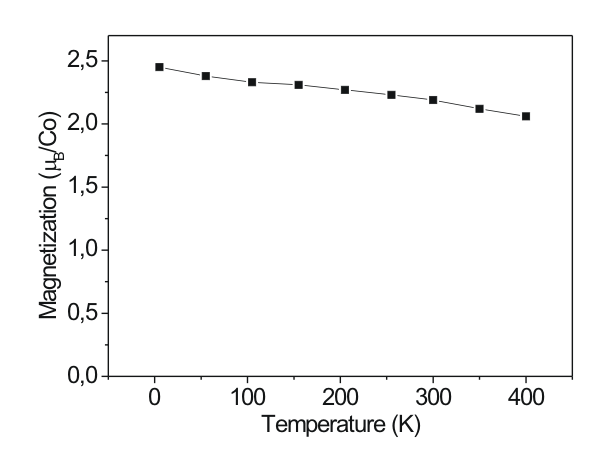}
\caption{\label{Squid-M-T-ZnO} Magnetization versus temperature
curve measured at H=5000 Oe by using a SQUID magnetometer. The solid
line is a guide to the eye.}
\end{figure}

\begin{figure}[!h]
\centering
\includegraphics[width=0.4\textwidth]{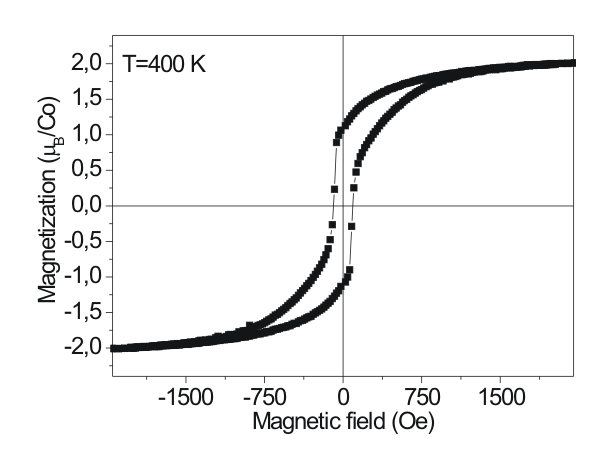}
\caption{\label{Squid-M-H-400K} Hysteresis curve measured at 400 K
by using a SQUID magnetometer.}
\end{figure}

In Fig.~\ref{Squid-M-T-ZnO} we show the temperature dependent
magnetization of Co-implanted ZnO film. Since the annealing to very
high temperatures destroys the ferromagnetism in oxide-based DMS
materials, \cite{KhaibullinNIMB07} we only heated the sample up to
400 K. From Fig.~\ref{Squid-M-T-ZnO} it is clear that the T$_C$ is
much higher than 400 K. To check whether the substitutional phase is
still present at 400 K, we measured another hysteresis curve at this
temperature using SQUID magnetometer. Fig.~\ref{Squid-M-H-400K}
shows that substitutional Co in ZnO is ferromagnetic even at 400 K.
By fitting a Brillouin curve to the M(T) data in
Fig.~\ref{Squid-M-T-ZnO} as a crude approximation, we estimate that
the ferromagnetic Curie temperature is as high as 700 K for our
film, as it was observed for Co-implanted TiO$_2$ rutile.
\cite{KhaibullinNIMB07}

\section{\label{sec:level1} Discussion and Conclusions}

In the literature, the reported highest solubility limit of cobalt
ions in ZnO is less than 50 \% using pulsed-laser deposition (PLD).
\cite{UedaAPL01} The measured cobalt concentration of 40-50 at.\% in
ZnO in this study is rather high, such that the formation of cobalt
clusters in ZnO should be expected. However, no clusters could be
observed within the ZnO layer. It seems that this is a peculiarity
of ion implantation, which allows doping of transition metals beyond
their solubility limits. \cite{HebardJPD04}

The difference in the shape of the hysteresis curves obtained by
MOKE and SQUID is attributed to the surface sensitivity of the MOKE
technique with a maximum penetration depth of about 20-30 nm. On the
other hand, the SQUID technique probes the whole volume of a sample.
The ZnO films has a thickness of 35 nm before implantation. Because
of the surface sputtering, the ZnO thickness decreases to about 28
nm after implantation. Thus MOKE provides information only from the
ZnO layer, not from the sapphire substrate, i.e. MOKE is only
sensitive to the ferromagnetic contribution from the ZnO layer. In
this layer a small fraction of nonmagnetic ZnO atoms are replaced by
magnetic Co ions, giving rise to the MOKE hysteresis. However, SQUID
collects magnetic contributions from both the Co-implanted ZnO film
and from the cobalt clusters in Al$_2$O$_3$. Therefore, the
difference between the MOKE and SQUID data appear as a result of the
depth-dependent Co content in the implanted layer.

Another important result of this study is the observation of oxygen
spin polarization in the Co-implanted ZnO film. Since the shape of
the hysteresis curve measured at the O \emph{K} edge
(Fig.~\ref{ZnO-Xrms-Hys-Co-O-S6}) is the same as the one recorded by
MOKE (Fig.~\ref{ZnO-Moke-Squid-Hys-S6}(a)), spin polarization of
oxygen atoms in this sample cannot be due to the cobalt clusters in
the sapphire substrate. Otherwise, the hysteretic shape of the
polarized oxygen should be similar to the hysteresis of metallic
cobalt clusters in sapphire with a large coercive field. From this
we infer that the oxygen atoms are polarized due to the spontaneous
ferromagnetic order in the ZnO film.

The main question that arises here is the mechanism which leads to
the observed long range ferromagnetic ordering in Co-doped ZnO.
Recently, Patterson \cite{PattersonCond-mat} calculated the
electronic structures of Co substituted for Zn in ZnO, for Zn and O
vacancies, and for interstitial Zn in ZnO using the B3LYP hybrid
density functional theory. He reported that the singly-positively
charged O vacancy is the only defect in Co-doped ZnO which can
mediate ferromagnetic exchange coupling of Co ions at intermediate
range (just beyond near neighbor distances). In the ground state
configuration the majority Co spins are parallel whereas the
minority spins are parallel to each other and to the oxygen vacancy
spin, so that there are exchange couplings between these three spins
which lead to an overall ferromagnetic ground state of the Co ions.
No substantial exchange coupling was found for the positively
charged interstitial Zn defect which has also spin half. The
exchange coupling mechanism explained by Patterson is essentially
the same as the impurity band model of Coey \emph{et al.},
\cite{CoeyNature05} in which the polarons bound to the oxygen
vacancies mediate ferromagnetic coupling between Co ions. In order
to have the magnetic moments of the Co ions aligned
ferromagnetically, one mediating electron is required with an
oppositely directed spin. The oxygen spin polarization has not
explicitly been considered in the aforementioned band structure
calculations and may be due to a ferromagnetic splitting of nearest
neighbor oxygen \emph{p}-levels.

\section{\label{sec:level1}Summary}

In conclusion, the structural and magnetic properties of a
Co-implanted ZnO film, deposited by RF-sputtering method on a
($11\overline{2}0$) oriented sapphire substrate, have been
investigated. The structural data indicate a Co cluster formation in
the sapphire substrate close to the ZnO/Al$_2$O$_3$ interface.
However, no indication of clustering in the ZnO layer has been
found. The XAS data with a multiplet structure around the Co
\emph{L}$_3$ edge clearly shows that the implanted cobalt ions are
in the Co$^{2+}$ oxidation state in Co-implanted ZnO film. The
magnetization measurements show that there are two magnetic phases
in the Co-implanted ZnO/Al$_2$O$_3$ films. One is the ferromagnetic
phase due to the Co substitution on Zn sites in the ZnO host matrix
and the second magnetic phase originates from Co clusters in the
sapphire substrate. Using x-ray resonant magnetic scattering at the
Co $\emph{L}_3$ edge, the magnetic contributions from the ZnO film
and the substrate can be separated. A clear ferromagnetic signal at
the O \emph{K} edge is also observed which shows that the oxygen
atoms close to the substituted cobalt atoms are polarized.
Furthermore, we have found very high magnetic magnetic moment of
2.81 $\mu_B$ per substituted cobalt atom with a very high Curie
temperature (T$_C$$\gg$400 K) in Co-implanted ZnO film.

\begin{acknowledgments}
We wish to acknowledge S. Erdt-B\"{o}hm and P. Stauche for sample
preparation and technical support and A. Kr\"{o}ger for preparation
of TEM samples. We would like to thank also Dr. C. Sternemann and
Dr. M. Paulus for their assistance with the beamline operation at
DELTA and G. Nowak for his help to perform XRD experiments at
HASYLAB. This work was partially supported by BMBF through Contracts
Nos. 05KS4PCA (ALICE Chamber) and 05ES3XBA/5 (Travel to BESSY), by
DFG through SFB 491, and by RFBR through the grant No 07-02-00559-a.
N. Akdogan acknowledges a fellowship through the IMPRS-SurMat.
\end{acknowledgments}

%\newpage %Just because of unusual number of tables stacked at end
\bibliography{ZnOpaper}% Produces the bibliography via BibTeX.

\end{document}